\documentclass[%
reprint,
superscriptaddress,
showpacs,preprintnumbers,
amsmath,amssymb,
prb,
]{revtex4-1}

\def\journal #1#2#3#4{#1 {\bf #2}, #3 (#4)}

\usepackage{graphicx}% Include figure files
\usepackage{dcolumn}% Align table columns on decimal point
\usepackage{bm}% bold math
\begin{document}

%\preprint{APS/123-QED}

\title{Strong coupling expansion in a correlated three-dimensional topological insulator}% Force line breaks with \\

\author{Akihiko Sekine}
\email{sekine@imr.tohoku.ac.jp}
\affiliation{Institute for Materials Research, Tohoku University, Sendai 980-8577, Japan}
\author{Takashi Z. Nakano}
\affiliation{Department of Physics and Yukawa Institute for Theoretical Physics, Kyoto University, Kyoto 606-8502, Japan}
\author{Yasufumi Araki}
\affiliation{Department of Physics, University of Texas at Austin, Austin, Texas 78712, USA}
\author{Kentaro Nomura}
\affiliation{Institute for Materials Research, Tohoku University, Sendai 980-8577, Japan}

\date{\today}% It is always \today, today,

\begin{abstract}
Motivated by recent studies which show that topological phases may emerge in strongly correlated electron systems, we theoretically study the strong electron correlation effect in a three-dimensional (3D) topological insulator, which effective Hamiltonian can be described by the Wilson fermions.
We adopt $1/r$ long-range Coulomb interaction as the interaction between the bulk electrons.
Based on the U(1) lattice gauge theory, the strong coupling expansion is applied by assuming that the effective interaction is strong.
It is shown that the effect of the Coulomb interaction is equivalent to the renormalization of the bare mass of the Wilson fermions, and that as a result, the topological insulator phase survives in the strong coupling limit.
\end{abstract}


\maketitle

%\tableofcontents
\section{Introduction}
Recently discovered topologically nontrivial phases have attracted many researchers and offered a new direction to modern physics\cite{Hasan2010,Qi2011}.
Topologically nontrivial phase and trivial phase, in the presence of time-reversal symmetry, are distinguished by the $Z_2$ invariant\cite{Fu2007,Fu2007a}.
Strong spin-orbit coupling is known to be essential to realize topological phases, since topological phases originate in the parity change in the lowest unoccupied band from even to odd induced by spin-orbit coupling.
Topological phases are characterized by the gapless edge (surface) states which are protected by time-reveral symmetry.
In 3D topological insulators, the surface states are described by the two-component massless Dirac fermions.
The bulk states in such as Bi$_2$Se$_3$ are described by the four-component anisotropic massive Dirac fermions\cite{Zhang2009}.
It is known that the surface states are robust against perturbation and disorder\cite{Bardarson2007,Nomura2007}.
What about against electron correlation,  i.e. Coulomb interaction?
This is a natural question, because it has been revealed that strong electron correlation is important in many systems and may induce novel phenomena.

A novel Mott-insulating phase was found recently in an iridate\cite{Kim2008}, a $5d$-electron system, and has gathered much attention.
Remarkably, the phase is induced by the cooperation of strong spin-orbit coupling and strong electron correlation.
Evolved by this discovery, many studies have been done intensively in systems where both spin-orbit coupling and electron correlation exist, for the search for novel phases induced by them.
Especially, it is of interest that topological phases such as the quantum spin Hall insulator\cite{Shitade2009} and the Weyl semimetal\cite{Wan2011} are predicted in iridates.
These results suggest that topological phases may emerge in strongly correlated $d$-electron systems.
Preceding studies mainly focus on the competition between the spin or charge ordered phase and the topological phase in Hubbard-like models on honeycomb lattices\cite{Raghu2008,Meng2010,Rachel2010,Varney2010,Hohenadler2011,Yamaji2011,Ruegg2011,Zheng2011,Yu2011}, other 2D lattices\cite{Sun2009,Wen2010,Yoshida2012,Yoshida2012a,Hohenadler2012} and 3D lattices\cite{Zhang2009a,Pesin2010,Mong2010,Kurita2011}.
Another study on the surface Dirac fermions shows that the Dirac fermions become massive with finite correlation strength due to the spotaneous magnetization\cite{Baum2012}.

On the other hand, the electron correlation effect in graphene, a two-dimensional Dirac fermion system, has been studied widely.
In graphene in vacuum, the coupling constant becomes effectively large due to the small Fermi velocity. It has been predicted that a finite band gap is induced in charge neutral graphene in vacuum.
In such a case, the strong coupling lattice gauge theory is applied\cite{Hands2008,Drut2009,Drut2009a,Armour2010,Drut2010,Araki2010,Araki2011,Araki2012,Buividovich2012}.
The chiral condensate is the order parameter for the insulator-semimetal transition in the lattice gauge theory.
It is noteworthy that lattice Monte Carlo studies show quantitatively correct critical value of the coupling strength below which the system becomes gapless\cite{Drut2009,Drut2009a,Buividovich2012} (graphene on a SiO$_2$ substrate is conducting).
These results motivated us to do this study.

In this paper, we focus on the strong electron correlation effect in a 3D Dirac fermion system on a lattice which is a simple model describing a topologically nontrivial state.
We adopt $1/r$ long-range Coulomb interaction as an interaction between the bulk electrons, because the screening effect in Dirac fermion systems is considered to be weak due to the vanishing of the density of states.
This situation is nothing but what is described by the U(1) lattice gauge theory.
Therefore, we can perform the strong coupling expansion of the lattice gauge theory by assuming that the effective coupling constant is large.
The procedure is as follows.
First we derive the effective action by the strong couling expansion.
Next we calculate the effective potential (the free energy per unit volume at zero temperature) with the use of the Hubbard-Stratonovich transformation and the mean-field approximation.
Finally we obtain the value of the chiral condensate as the stationary point of the effective potential.
Our model, the Wilson fermions, breaks chiral symmetry by itself, and thus we cannot use the chiral condensate as the order parameter for the the insulator-semimetal transition.
We regard the chiral condensate as a correction to the bare mass.

The main purpose of this study is devided into two parts:
(I) answer the question that whether the topological insulator phase survives at the limit of infinitely strong Coulomb interaction between the bulk electrons, or not. 
To do this, we have to obtain the value of the chiral condensate, which corresponds to a correction to the bare mass, in the strong coupling limit.
(II) search for the phase in which time-reversal and inversion symmetries are spontaneously broken due to electron correlation.
Such a phase, "Aoki phase" has been confirmed in the lattice quantum chromodynamics (QCD) with Wilson fermions\cite{Aoki1984,Aoki1986,Sharpe1998} and was suggested recently in a mean-field study of Wilson fermions with the short-range interaction\cite{Sekine2012}.

\section{Model}
It is known that the effective Hamiltonian of 3D topological insulators such as Bi$_2$Se$_3$ is described by the Wilson fermion\cite{Zhang2009}:
\begin{equation}
\begin{aligned}
\mathcal{H}_{0}(\bm{k})={\sum}_j\sin k_j\cdot\alpha_j+m(\bm{k})\beta,\label{H_0}
\end{aligned}
\end{equation}
where $m(\bm{k})=m_0+r\sum_j\left(1-\cos k_j\right)$, $r>0$, $j\ (=1,2,3)$ denotes spacial axis, and $\alpha_j$, $\beta$ are the Dirac gamma matrices given by
\begin{equation}
\begin{aligned}
\alpha_j=
\begin{bmatrix}
0 & \sigma_j\\
\sigma_j & 0
\end{bmatrix},\ \ \ \ \
\beta=
\begin{bmatrix}
1 & 0\\
0 & -1
\end{bmatrix}.
\end{aligned}
\end{equation}
The energy of this system is measured in units of $v_{\mathrm{F}}/a$ with $v_{\mathrm{F}}$ and $a$ being the Fermi velocity and the lattice constant, respectively.
The Hamiltonian (\ref{H_0}) has time-reversal ($\mathcal{T}$) symmetry and inversion ($\mathcal{I}$) symmetry, i.e., $\mathcal{T}\mathcal{H}_{0}(\bm{k})\mathcal{T}^{-1}=\mathcal{H}_{0}(-\bm{k})$ and $\mathcal{I}\mathcal{H}_{0}(\bm{k})\mathcal{I}^{-1}=\mathcal{H}_{0}(-\bm{k})$ are satisfied, where $\mathcal{T}=\bm{1}\otimes(-i\sigma_2)\mathcal{K}$ ($\mathcal{K}$ is the complex conjugation operator) and $\mathcal{I}=\sigma_3\otimes\bm{1}$.
In the Hamiltonian (\ref{H_0}), the spinor is written in the basis of $\left[c^\dag_{\bm{k}A\uparrow},c^\dag_{\bm{k}A\downarrow},c^\dag_{\bm{k}B\uparrow},c^\dag_{\bm{k}B\downarrow}\right]$, where $c^\dag$ is the creation operator of an electron, $A$, $B$ denote two orbitals, and $\uparrow$ ($\downarrow$) denotes up- (down-) spin\cite{Zhang2009}.

In the presence of time-reversal symmetry and inversion symmetry, the $Z_2$ invariant of the system is given by\cite{Fu2007,Fu2007a}
\begin{equation}
\begin{aligned}
(-1)^\nu=\prod_{i=1}^8\left\{-\mathrm{sgn}\left[m\left(\bm{\Lambda}_i\right)\right]\right\}, \label{Z2invariant}
\end{aligned}
\end{equation}
where $\bm{\Lambda}_i$ are the eight time-reversal invariant momenta.
It is easily shown that if $0>m_0>-2r$ or $-4r>m_0>-6r$ ($m_0>0$, $-2r>m_0>-4r$, or $-6r>m_0$), the system is topologically nontrivial (trivial).

Let us consider a strongly correlated topological insulator in the Euclidean spacetime, which is described by the Wilson fermions with $1/r$ Coulomb interaction between the bulk electrons.
We start from the Euclidean action of (3+1)D Wilson fermion interacting with electromagnetic field on a lattice, which is given by
\begin{equation}
\begin{aligned}
S_{F}=&-\sum_{n,\mu}\left[\bar{\psi}_nP^-_\mu U_{n,\mu}\psi_{n+\hat{\mu}} + \bar{\psi}_{n+\hat{\mu}}P^+_\mu U^\dag_{n,\mu}\psi_n\right]\\
&+(m_0+4r)\sum_{n}\bar{\psi}_n \psi_n,\label{ActionQED}
\end{aligned}
\end{equation}
where $P^\pm_\mu=(r\pm \gamma_\mu)/2$.
Here $n=(n_0,n_1,n_2,n_3)$ denotes a site on a spacetime lattice and $\hat{\mu}$ ($\mu=0,1,2,3$) denotes the unit vector along $\mu$-direction.
$U_{n,\mu}$ is the link variable, which is defined by $U_{n,\mu}=e^{iagA_{\mu}(n+\hat{\mu}/2)}$, where $A_\mu=(A_0,\bm{A})$ is the four-vector potential, $a$ is the lattice constant, and $g^2=e^2/\epsilon$ with $e$ and $\epsilon$ being electric charge and the permittivity of the system, respectively.
Although the timelike Wilson term (the term proportional to $r$) is introduced artificially to eliminate fermion doublers, the spatial Wilson terms have a physical meaning (arise due to strong spin-orbit coupling).
In this paper, according to the Hamiltonian (\ref{H_0}), we adopt the Dirac representation in the Euclidean spacetime ($\{\gamma_\mu,\gamma_\nu\}=2\delta_{\mu\nu}$):
\begin{equation}
\begin{aligned}
\gamma_0=
\begin{bmatrix}
1 & 0\\
0 & -1
\end{bmatrix},\ \ \ 
\gamma_j=
\begin{bmatrix}
0 & -i\sigma_j\\
i\sigma_j & 0
\end{bmatrix},\ \ \ 
\gamma_5=
\begin{bmatrix}
0 & 1\\
1 & 0
\end{bmatrix},
\end{aligned}\label{gamma-matrices}
\end{equation}
where $j=1,2,3$ and $\sigma_j$ are the Pauli matrices.

In the case of 3D topological insulators, the Fermi velocity $v_{\rm F}$ is about $3\times 10^{-3}c$ where $c$ is the speed of light in vacuum.
Then the interactions between the bulk electrons can be regarded as only the instantaneous Coulomb interaction ($A_j=0$) like in the case of graphene\cite{Hands2008,Drut2009,Drut2009a,Armour2010,Drut2010,Araki2010,Araki2011,Araki2012,Buividovich2012}, so the action (\ref{ActionQED}) is rewritten as
\begin{equation}
\begin{aligned}
S_{F}=S_F^{(\tau)}+S_F^{(s)}+(m_0+4r)\sum_{n}\bar{\psi}_n \psi_n,\label{ActionTI}
\end{aligned}
\end{equation}
where
\begin{equation}
\left\{
\begin{aligned}
S_F^{(\tau)}&=-\sum_{n}\left[\bar{\psi}_nP^-_0 U_{n,0}\psi_{n+\hat{0}} + \bar{\psi}_{n+\hat{0}}P^+_0 U^\dag_{n,0}\psi_n\right]\\
S_F^{(s)}&=-\sum_{n,j}\left[\bar{\psi}_nP^-_j\psi_{n+\hat{j}} + \bar{\psi}_{n+\hat{j}}P^+_j\psi_n\right],
\end{aligned}
\right.
\end{equation}
and $U_{n,0}=e^{i\theta_n}\ (-\pi\leq\theta_n\leq\pi)$.
The Wilson fermions breaks chiral symmetry by itself (the terms proportional to $r$ and $m_0$), i.e., the action (\ref{ActionTI}) is not invariant under the chiral transformation $\psi\rightarrow e^{i\theta\gamma_5}\psi$.
In our model, chiral symmetry is equivalent to the symmetry of the pseudospin for two $p$-orbitals $A$ and $B$.
The pure U(1) gauge action on a lattice is given by
\begin{equation}
\begin{aligned}
S_G=\beta\sum_n\sum_{\mu>\nu}\left[1-\frac{1}{2}\left(U_{n,\mu\nu}+U^\dag_{n,\mu\nu}\right)\right],
\end{aligned}
\end{equation}
where $\beta=v_{\rm F}/g^2$. The plaquette contribution $U_{n,\mu\nu}$ is defined by
\begin{equation}
\begin{aligned}
U_{n,\mu\nu}=U_{n,\mu}U_{n+\hat{\mu},\nu}U^\dag_{n+\hat{\nu},\mu}U^\dag_{n,\nu},
\end{aligned}
\end{equation}
where $U_{n,j}=1$ in our case.
The total action on a lattice is written as
\begin{equation}
\begin{aligned}
S=S_F+S_G.
\end{aligned}
\end{equation}
The dielectric constant $\epsilon_r$ of Bi$_2$Se$_3$ is rather large\cite{dielectricconstant-Bi2Se3} ($\epsilon_r=\epsilon/\epsilon_0\approx 100$).
This means that the Coulomb interaction between the bulk electrons in Bi$_2$Se$_3$ is considered to be weak.
In fact, the value of $\beta$ is approximated as
\begin{equation}
\begin{aligned}
\beta=\frac{v_{\rm F}\epsilon_r}{4\pi c}\cdot\frac{4\pi\epsilon_0\hbar c}{e^2}\approx 3,
\end{aligned}
\end{equation}
and we cannot perform the strong coupling expansion in Bi$_2$Se$_3$.
However, we think it would be important from a theorerical viewpoint to examine the strong electron correlation effect in Dirac fermion systems which describe topologically nontrivial states.

\section{Effective Action}
Let us perform the strong coupling expansion.
The strong coupling expansion has  been often used in QCD\cite{Kawamoto1981,Hoek1982,Drouffe1983,Nishida2004,Miura2009} where the coupling between fermions (quarks) and gauge fields (gluons) are strong.
We can carry out the $U_0$ integral by using the SU($N_c$) group integral formulae:
\begin{equation}
\begin{aligned}
\int dU1=1,\ \ \int dUU_{ab}=0,\ \ \int dUU_{ab}U_{cd}^\dag=\frac{1}{N_c}\delta_{ad}\delta_{bc}.
\end{aligned}
\end{equation}
Our case corresponds to the case of $N_c=1$.
In the following, we derive the effective action $S_{\rm eff}[\psi,\bar{\psi}]$ by carrying out  the $U_0$ integral:
\begin{equation}
\begin{aligned}
Z=\int \mathcal{D}[\psi,\bar{\psi},U_0]e^{-S_F-S_G}=\int \mathcal{D}[\psi,\bar{\psi}]e^{-S_{\rm eff}}.
\end{aligned}
\end{equation}

First we consider the strong coupling limit ($\beta=0$). In this case, the timelike partition function is given by
\begin{equation}
\begin{aligned}
Z^{(\tau)}_{\mathrm{SCL}}[\psi,\bar{\psi}]=\int \mathcal{D}U_0e^{-S^{(\tau)}_F}.
\end{aligned}
\end{equation}
Integration with respect to $U_0$ is carried out to be
\begin{equation}
\begin{aligned}
&Z^{(\tau)}_{\mathrm{SCL}}=\exp\left[\sum_n\bar{\psi}_nP^-_0 \psi_{n+\hat{0}}\bar{\psi}_{n+\hat{0}}P^+_0 \psi_n\right].\label{Z_SCL}
\end{aligned}
\end{equation}
Here we have used the fact that the grassmann variables $\psi$'s and $\bar{\psi}$'s satisfy $\psi^2=\bar{\psi}^2=0$. We can rewrite this term as
\begin{equation}
\begin{aligned}
\bar{\psi}_nP^-_0 \psi_{n+\hat{0}}\bar{\psi}_{n+\hat{0}}P^+_0 \psi_n= -\mathrm{tr}\left[M_nP^+_0M_{n+\hat{0}}P^-_0\right],
\end{aligned}
\end{equation}
where we have defined $(M_n)_{\alpha\beta}=\bar{\psi}_{n,\alpha}\psi_{n,\beta}$ and used $(P^\pm_0)_{\alpha\beta}=(P^\pm_0)_{\beta\alpha}$. The subscripts $\alpha$ and $\beta$ denote the component of spinors.

Next we evaluate the term of the order of $\beta$. 
In order to evaluate the plaquette contributions from $S_G$, we use the cumulant expansion\cite{Miura2009,Kubo1962}.
Let us define an expectation value:
\begin{equation}
\begin{aligned}
\left\langle A\right\rangle&\equiv \frac{1}{Z^{(\tau)}_{\mathrm{SCL}}}\int \mathcal{D}U_0A[U_0]e^{-S^{(\tau)}_F}.
\end{aligned}
\end{equation}
Then using this definition, the full timelike partition function can be expressed as 
\begin{equation}
\begin{aligned}
Z^{(\tau)}=\int \mathcal{D}U_0e^{-S^{(\tau)}_F-S_G}=Z^{(\tau)}_{\mathrm{SCL}}\left\langle e^{-S_G}\right\rangle.\label{Z-tau-full}
\end{aligned}
\end{equation}
The contribution from $S_G$ is given by
\begin{equation}
\begin{aligned}
\Delta S\equiv -\log\left\langle e^{-S_G}\right\rangle=-\sum_{n=1}^{\infty}\frac{(-1)^n}{n!}\left\langle S_G^n\right\rangle_c,
\end{aligned}
\end{equation}
where $\left\langle \cdots\right\rangle_c$ is a cumulant.
The correction to the action up to $\mathcal{O}(\beta)$ is given by
\begin{equation}
\begin{aligned}
\Delta S&=\left\langle S_G\right\rangle_c=\left\langle S_G\right\rangle\\
&=-\frac{\beta}{2}\sum_n\sum_{\mu>\nu}\left\langle U_{n,\mu\nu}+U^\dag_{n,\mu\nu}\right\rangle.\label{S-NLO}
\end{aligned}
\end{equation}
The expectation value of $U_{n,\mu\nu}$ is evaluated as follows\cite{Miura2009}:
\begin{equation}
\begin{aligned}
\left\langle U_{n,\mu\nu}\right\rangle\simeq \int dU_{n,0} U_{n,\mu\nu}e^{-s^{(\tau)}_P},
\end{aligned}
\end{equation}
where $s^{(\tau)}_P$ is the plaquette-related part of $S^{(\tau)}_F$.
We see that the terms with $(\mu,\nu)=(i,j)$ become constant and find only $(\mu,\nu)=(j,0)$ terms to survive:
\begin{equation}
\left\{
\begin{aligned}
\left\langle U_{n,j0}\right\rangle&=-\mathrm{tr}\left[V^+_{n,j}P^+_0V^-_{n+\hat{0},j}P^-_0\right],\\
\left\langle U^\dag_{n,j0}\right\rangle&=-\mathrm{tr}\left[V^-_{n,j}P^+_0V^+_{n+\hat{0},j}P^-_0\right],
\end{aligned}
\right.
\end{equation}
where we have defined $\left(V^+_{n,j}\right)_{\alpha\beta}=\bar{\psi}_{n,\alpha}\psi_{n+\hat{j},\beta}$ and $\left(V^-_{n,j}\right)_{\alpha\beta}=\bar{\psi}_{n+\hat{j},\alpha}\psi_{n,\beta}$.

Finally, substituting Eqs. (\ref{Z_SCL}) and (\ref{S-NLO}) to Eq. (\ref{Z-tau-full}), we obtain the effective action up to $\mathcal{O}(\beta)$:
\begin{equation}
\begin{aligned}
S_{\mathrm{eff}}=&(m_0+4r)\sum_{n}\bar{\psi}_n\psi_n-\sum_{n,j}\left[\bar{\psi}_nP^-_j\psi_{n+\hat{j}} + \bar{\psi}_{n+\hat{j}}P^+_j\psi_n\right]\\
&+\sum_n\mathrm{tr}\left[M_nP^+_0M_{n+\hat{0}}P^-_0\right]\\
&+\frac{\beta}{2}\sum_{n,j}\left\{\mathrm{tr}\left[V^+_{n,j}P^+_0V^-_{n+\hat{0},j}P^-_0\right]+(V^+\longleftrightarrow V^-)\right\}.\label{Eff-Act}
\end{aligned}
\end{equation}

\section{Effective Potential and Chiral Condensate}
In this section, we derive the effective potential with the use of the extended Hubbard-Stratonovich transformation (EHS)\cite{Miura2009,Araki2010,Araki2011,Araki2012}, and then we obtain the value of the chiral condensate as the stationary point of the effective potential.
We apply the EHS to the trace of arbitrary two matrices. Introducing two auxiliary fields $R$ and $R'$, we obtain
\begin{equation}
\begin{aligned}
&e^{\kappa\mathrm{tr}AB}\propto\\
&\hspace{0.3cm}\int \mathcal{D}[R,R'] \exp\left\{-\kappa\sum_{\alpha\beta}\left[(R_{\alpha\beta})^2+(R'_{\alpha\beta})^2\rule{0pt}{3ex}\right.\right.\\
&\hspace{0.5cm}\left.\left.-(A_{\alpha\beta}+B^T_{\alpha\beta})R_{\beta\alpha}-i(A_{\alpha\beta}-B^T_{\alpha\beta})R'_{\beta\alpha}\rule{0pt}{3ex}\right]\rule{0pt}{5ex}\right\},\label{EHS}
\end{aligned}
\end{equation}
where $\kappa$ is a positive constant and the superscript $T$ denotes the transpose of a matrix.
Two auxiliary fields take the saddle point values $R_{\alpha\beta}=\left\langle A+B^T\right\rangle_{\beta\alpha}$/2 and $R'_{\alpha\beta}=i\left\langle A-B^T\right\rangle_{\beta\alpha}$/2, respectively. 
Defining $Q=R+iR'$ and $Q'=R-iR'$, Eq. (\ref{EHS}) is rewritten as
\begin{equation}
\begin{aligned}
&e^{\kappa\mathrm{tr}AB}\propto\\
&\int \mathcal{D}[Q,Q'] \exp\left\{-\kappa\left[Q_{\alpha\beta}Q'_{\alpha\beta}-A_{\alpha\beta}Q_{\beta\alpha}-B^T_{\alpha\beta}Q'_{\beta\alpha}\right]\right\},\label{EHS-Q}
\end{aligned}
\end{equation}
with the saddle point values $Q_{\alpha\beta}=\left\langle B^T\right\rangle_{\beta\alpha}$ and $Q'_{\alpha\beta}=\left\langle A\right\rangle_{\beta\alpha}$.

\subsection{Effective Potential in the Strong Coupling Limit}
We consider to decouple the third term in the effective action (\ref{Eff-Act}) to fermion bilinear form.
To do this, we set $(\kappa, A, B)=(1, M_nP^+_0, -M_{n+\hat{0}}P^-_0)$ in Eq. (\ref{EHS-Q}). 
In this case, the saddle point values are given by $Q_{\alpha\beta}=-\left\langle M_{n+\hat{0}}P^-_0\right\rangle_{\alpha\beta}$ and $Q'_{\alpha\beta}=\left\langle M_nP^+_0\right\rangle_{\beta\alpha}$.
Here let us assume that
\begin{equation}
\begin{aligned}
\left\langle M_n\right\rangle&=\sigma e^{i\theta\gamma_5}=\sigma (\cos\theta I+i\sin\theta\gamma_5)\\
&=\sigma
\begin{bmatrix}
\cos\theta & i\sin\theta\\
i\sin\theta & \cos\theta
\end{bmatrix},\label{M_n}
\end{aligned}
\end{equation}
because we are now interested in the phase structure of the Wilson fermions interacting via the long-range Coulomb interaction in the strong coupling limit, i.e., the mass term (the terms proportional to the identity matrix) is important when determining the phase is whether  topologically trivial or nontrivial (see Eq. (\ref{Z2invariant})).
We are also interested in the possibility of the existence of the symmetry broken phase ("Aoki phase") in this model.
Thus the pseudoscalar modes $i\gamma_5$ should be taken into account.
Then it follows that
\begin{equation}
\left\{
\begin{aligned}
\left\langle \bar{\psi}\psi\right\rangle&=\sigma\cos\theta\equiv \phi_\sigma\\
\left\langle \bar{\psi}i\gamma_5\psi\right\rangle&=\sigma\sin\theta\equiv \phi_\pi.
\end{aligned}
\right.
\end{equation}
The terms $\left\langle \bar{\psi}\psi\right\rangle$ and $\left\langle \bar{\psi}i\gamma_5\psi\right\rangle$ decribe the chiral condensate and the condensate of pseudoscalar mode, respectively.

The Wilson fermions breaks chiral symmetry by itself (the terms proportional to $r$ and $m_0$).
Hence we cannot use the value of $\left\langle \bar{\psi}\psi\right\rangle$ to determine the system is whether insulating or semimetallic, unlike in the case of graphene where chiral symmetry is not broken in the noninteracting limit. 
We regard the value of the chiral condensate $\left\langle \bar{\psi}\psi\right\rangle$ as a correction to the bare mass.

Substituting $(\kappa, A, B)=(1, M_nP^+_0, -M_{n+\hat{0}}P^-_0)$ to Eq. (\ref{EHS-Q}), we obtain
\begin{equation}
\begin{aligned}
&\exp\left\{-\sum_n\mathrm{tr}\left[M_nP^+_0M_{n+\hat{0}}P^-_0\right]\right\}\\
&\sim\exp\left\{-\sum_{n}\left[(1-r^2)\phi_\sigma^2+(1+r^2)\phi_\pi^2\rule{0pt}{3ex}\right.\right.\\
&\left.\left.\hspace{0.35cm}+\frac{1}{2}\bar{\psi}_n\left[-(1-r^2)\phi_\sigma+i\gamma_5^T(1+r^2)\phi_\pi\right]\psi_n\rule{0pt}{3ex}\right]\rule{0pt}{4ex}\right\},
\end{aligned}
\end{equation}
where we have applied the mean-field approximation for the chiral condensate and the condensate of pseudoscalar mode.
Thus the effective action in the strong coupling limit expressed by the two auxiliary fields $\phi_\sigma$ and $\phi_\pi$ is given by
\begin{equation}
\begin{aligned}
S_{\mathrm{eff}}(\phi_\sigma,\phi_\pi)=&N_sN_\tau\left[(1-r^2)\phi_\sigma^2+(1+r^2)\phi_\pi^2\right]\\
&+\sum_k \bar{\psi}_k\mathcal{M}(\bm{k};\phi_\sigma,\phi_\pi)\psi_k, \label{S_eff_SCL}
\end{aligned}
\end{equation}
with
\begin{equation}
\begin{aligned}
\mathcal{M}=&{\sum}_j i\gamma_j\sin k_j+m_0+r\left(4-{\sum}_j\cos k_j\right)\\
&-\frac{1}{2}(1-r^2)\phi_\sigma+i\gamma_5^T\frac{1}{2}(1+r^2)\phi_\pi.\label{M-SCL}
\end{aligned}
\end{equation}
Here $N_s=V$ and $N_\tau=1/T$ with $V$ and $T$ being the volume and the temperature of the system, respectively and we have done the Fourier transform from $n=(n_0,\bm{n})$ to $k=(k_0,\bm{k})$.

The effective potential at zero temperature per unit spacetime volume is given by
\begin{equation}
\begin{aligned}
\mathcal{F}_{\mathrm{eff}}(\phi_\sigma,\phi_\pi)=-\frac{1}{N_sN_\tau}\log Z(\phi_\sigma,\phi_\pi).
\end{aligned}
\end{equation}
Integration with respect to $\psi$ and $\bar{\psi}$ is carried out by the formula $\int D[\psi,\bar{\psi}]e^{-\bar{\psi}\mathcal{M}\psi}=\mathrm{det}\mathcal{M}$. 
Therefore we need to calculate the determinant of $\mathcal{M}$.
From Eq. (\ref{M-SCL}), the matrix $\mathcal{M}$ is written explicitly as
\begin{equation}
\begin{aligned}
\mathcal{M}&=
\begin{bmatrix}
\tilde{m}(\bm{k})+r & \sigma_j\sin k_j+i\frac{1+r^2}{2}\phi_\pi\\
-\sigma_j\sin k_j+i\frac{1+r^2}{2}\phi_\pi & \tilde{m}(\bm{k})+r
\end{bmatrix}\\
&\equiv
\begin{bmatrix}
A & B\\
C & D
\end{bmatrix},
\end{aligned}\label{M}
\end{equation}
where
\begin{equation}
\begin{aligned}
\tilde{m}(\bm{k})=m_0-\frac{1-r^2}{2}\phi_\sigma+r{\sum}_j\left(1-\cos k_j\right). \label{m_eff}
\end{aligned}
\end{equation}
As we see from Eq. (\ref{m_eff}), the chiral condensate $\phi_\sigma$ corresponds to a correction to the bare mass $m_0$ in the original Hamiltonian (\ref{H_0}).
That is, $m(\bm{k})$ in the noninteracting Hamiltonian (\ref{H_0}) changes to $\tilde{m}(\bm{k})$ in the strong coupling limit.
The term "$r$" of $\tilde{m}(\bm{k})+r$ in Eq. (\ref{M}) originates in the timelike components of the action.
After a straightforward calculation, we have
\begin{equation}
\begin{aligned}
&\mathrm{det}\mathcal{M}=\mathrm{det}A\cdot\mathrm{det}\left(D-CA^{-1}B\right)\\
&=\left[{\sum}_j\sin^2k_j+\left[\tilde{m}(\bm{k})+r\right]^2+\frac{(1+r^2)^2}{4}\phi_\pi^2\right]^2.\label{detM}
\end{aligned}
\end{equation}
The same result can be derived by the formula $\mathrm{det}\mathcal{M}=\sqrt{\mathrm{det}(\mathcal{M}\mathcal{M}^\dag)}$.
Finally we arrive at the effective potential in the strong coupling limit:
\begin{equation}
\begin{aligned}
&\mathcal{F}_{\mathrm{eff}}(\phi_\sigma,\phi_\pi)=(1-r^2)\phi_\sigma^2+(1+r^2)\phi_\pi^2-2\int_{-\pi}^{\pi}\frac{d^3k}{(2\pi)^3}\\
&\times\log\left[{\sum}_j\sin^2k_j+\left[\tilde{m}(\bm{k})+r\right]^2+\frac{(1+r^2)^2}{4}\phi_\pi^2\right].\label{Feff-SCL}
\end{aligned}
\end{equation}

The values of $\phi_\sigma$ and $\phi_\pi$ are obtained by the stationary conditions $\partial \mathcal{F}_{\mathrm{eff}}(\phi_\sigma,\phi_\pi)/\partial \phi_\sigma=\partial \mathcal{F}_{\mathrm{eff}}(\phi_\sigma,\phi_\pi)/\partial \phi_\pi=0$.
When $r=1$, Eq. (\ref{Feff-SCL}) does not depend on $\phi_\sigma$.
In this case,  the stationary point is obtained by the following equation:
\begin{equation}
\begin{aligned}
\phi_\sigma&=\frac{1}{4N_sN_\tau}\frac{\int \mathcal{D}[\psi,\bar{\psi},U_0]\sum_n\bar{\psi}_n\psi_n e^{-S}}{\int \mathcal{D}[\psi,\bar{\psi},U_0]e^{-S}}\\
&=-\frac{1}{4N_sN_\tau}\frac{1}{Z}\frac{d Z}{d m_0}\\
&=\frac{1}{4}\frac{d\mathcal{F}_{\mathrm{eff}}}{d m_0}.
\end{aligned}
\end{equation}
When $r>1$, the coefficient of the first term in Eq. (\ref{Feff-SCL}), $1-r^2$, becomes negative and thus Eq. (\ref{Feff-SCL}) does not have the stationary point.
This is because the logarithmic term is doninant when $\phi_\sigma$ is small and then $\phi_\sigma^2$ term becomes dominant as $\phi_\sigma$ gets larger.
Therefore the condition that the coefficient of $\phi_\sigma^2$ must be positive is needed for Eq. (\ref{Feff-SCL}) to have the stationary point.
This fact is consistent with the requirement of the reflection positivity of lattice gauge theories with Wilson fermions\cite{Menotti1987}.

In the chiral limit ($r=m_0=0$), the effective potential is a function of only $\sigma$, reflecting the chiral symmetry of the action.
This is understood as follows: in the chiral limit, the action is invariant under the chiral transformation $\psi\rightarrow e^{i\theta\gamma_5}\psi$.
This transformation doesn't depend on the value of $\theta$, and thus the effective potential also doesn't depend on it.
Note that this effective potential corresponds to that of the staggered fermion (SF) model for graphene\cite{Araki2010,Araki2011} except for an additional factor 4 by setting $r=0$ and changing from (3+1)D to (2+1)D:
\begin{equation}
\begin{aligned}
\mathcal{F}_{\mathrm{eff}}(\phi_\sigma,\phi_\pi)=4\mathcal{F}^{\mathrm{SF}}_{\mathrm{eff}}(\phi_\sigma,\phi_\pi).
\end{aligned}
\end{equation}
This result is reasonable, because the two cases describes the same system where the $2^3=8$ fermion doublers appear.

\subsection{Effective Potential Up to $\bm{\mathcal{O}(\beta)}$}
Let us evaluate the $\mathcal{O}(\beta)$ contribution to the effective potential.
We write the fourth term in the effective action (\ref{Eff-Act}) as $\Delta S_1+\Delta S_2(\equiv \Delta S)$.
Then we should choose such that $(\kappa, A, B)=(\beta/2, V^+_{n,j}P^+_0, -V^-_{n+0,j}P^-_0)$ in Eq. (\ref{EHS-Q}) for $\Delta S_1$:
\begin{equation}
\begin{aligned}
e^{-\Delta S_1}
\propto&\exp\left\{-\frac{\beta}{2}\sum_{n,j}\left[S_{\alpha\beta}S'_{\alpha\beta}-A_{\alpha\beta}S_{\beta\alpha}-B^T_{\alpha\beta}S'_{\beta\alpha}\right]\right\}, \label{Delta S_1}
\end{aligned}
\end{equation}
with the saddle point values $S_{\alpha\beta}=\left\langle B^T\right\rangle_{\beta\alpha}=-\left\langle V^-_{n+0,j}P^-_0\right\rangle_{\alpha\beta}$ and $S'_{\alpha\beta}=\left\langle A\right\rangle_{\beta\alpha}=\left\langle V^+_{n,j}P^+_0\right\rangle_{\beta\alpha}$.
Similarly, setting $(\kappa, A, B)=(\beta/2, V^-_{n,j}P^+_0, -V^+_{n+0,j}P^-_0)$ in Eq. (\ref{EHS-Q}) for $\Delta S_2$, we obtain
\begin{equation}
\begin{aligned}
e^{-\Delta S_2}
\propto&\exp\left\{-\frac{\beta}{2}\sum_{n,j}\left[T_{\alpha\beta}T'_{\alpha\beta}-A_{\alpha\beta}T_{\beta\alpha}-B^T_{\alpha\beta}T'_{\beta\alpha}\right]\right\}, \label{Delta S_2}
\end{aligned}
\end{equation}
with the saddle point values $T_{\alpha\beta}=\left\langle B^T\right\rangle_{\beta\alpha}=-\left\langle V^+_{n+0,j}P^-_0\right\rangle_{\alpha\beta}$ and $T'_{\alpha\beta}=\left\langle A\right\rangle_{\beta\alpha}=\left\langle V^-_{n,j}P^+_0\right\rangle_{\beta\alpha}$.

Next we decompose $\left\langle V^+_{n,j}\right\rangle$ and $\left\langle V^-_{n,j}\right\rangle$ into spinor components as follows:
\begin{equation}
\left\{
\begin{aligned}
\langle V_{n,j}^+ \rangle & \equiv v_s^+ + i \gamma_5 v_p^+ + \sum_\mu \gamma_\mu v_{v \mu}^+ + \sum_\mu i \gamma_5 \gamma_\mu v_{a \mu}^+,\\
\langle V_{n,j}^- \rangle & \equiv v_s^- + i \gamma_5 v_p^- + \sum_\mu \gamma_\mu v_{v \mu}^- + \sum_\mu i \gamma_5 \gamma_\mu v_{a \mu}^-,
\end{aligned}
\right. \label{V^+-V^-}
\end{equation}
where the first, second, third and fourth terms are the components of scalar, pseudoscalar, vector and pseudovector (axial vector) mode, respectively.
The terms $\left\langle V^+_{n,j}\right\rangle$ and $\left\langle V^-_{n,j}\right\rangle$ are equivalent to the propagator from a point to another point.
Only the scalar and vector modes appear when parity is not broken, and the pseudoscalar and pseudovector modes may also appear when parity is broken.
Therefore these four modes should be considered in Eq. (\ref{V^+-V^-}).

After the calculation in the appexdix, we obtain the $\mathcal{O}(\beta)$ contribution to the action as
\begin{widetext}
\begin{equation}
\begin{aligned}
\Delta S=&\beta\sum_{n,j}\left[(1-r^2) v_s^- v_s^+ +(1+r^2) v_p^- v_p^+ +(1-r^2) v_{v0}^- v_{v0}^+ - (1+r^2) \sum_l v_{vl}^- v_{vl}^+ -(1+r^2) v_{a0}^- v_{a0}^+ + (1-r^2) \sum_l v_{al}^- v_{al}^+ \right]\\
&+\sum_{n,j} \left[ \bar{\psi}_n \mathcal{A}_- \psi_{n+\hat{j}} + \bar{\psi}_{n+\hat{j}} \mathcal{A}_+ \psi_n \right], \label{Delta S}
\end{aligned}
\end{equation}
where
\begin{equation}
\begin{aligned}
\langle \mathcal{A}_- \rangle &=\frac {\beta}{4}\left[
-(1-r^2) v_s^- + (1+r^2) i \gamma_5 v_p^- - (1-r^2)\gamma_0 v_{v0}^- 
+ (1+r^2)\sum_l \gamma_l  v_{vl}^- +  (1+r^2)i \gamma_5 \gamma_0 v_{a0}^- - (1-r^2)\sum_l i \gamma_5 \gamma_l v_{al}^-\right]^T,\\
\langle \mathcal{A}_+ \rangle &=\frac {\beta}{4}\left[
-(1-r^2) v_s^+ + (1+r^2) i \gamma_5 v_p^+ - (1-r^2)\gamma_0 v_{v0}^+ 
+ (1+r^2)\sum_l \gamma_l  v_{vl}^+ +  (1+r^2)i \gamma_5 \gamma_0 v_{a0}^+ - (1-r^2)\sum_l i \gamma_5 \gamma_l v_{al}^+\right]^T.
\end{aligned}
\end{equation}
Then doing the Fourier transform and combining Eqs. (\ref{S_eff_SCL}) and (\ref{Delta S}), we get the effective action up to $\mathcal{O}(\beta)$ with auxiliary fields:
\begin{equation}
\begin{aligned}
S_{\mathrm{eff}}=S_{\mathrm{eff}}^{\mathrm{aux}}\left(\phi_\sigma,\phi_\pi, v^\pm_s, v^\pm_p, v^\pm_{v\mu}, v^\pm_{a\mu}\right)+\sum_k \bar{\psi}_k\mathcal{M}\left(\bm{k};\phi_\sigma,\phi_\pi, v^\pm_s, v^\pm_p, v^\pm_{v\mu}, v^\pm_{a\mu}\right)\psi_k.
\end{aligned}
\end{equation}
For the explicit forms of $S_{\mathrm{eff}}^{\mathrm{aux}}$ and $\mathcal{M}$, see the appendix.

Finally, after eliminating the auxiliary fields $v$'s by the stationary conditions, we arrive at the effective potential up to $\mathcal{O}(\beta)$ given by
\begin{equation}
\begin{aligned}
\mathcal{F}_{\mathrm{eff}}\left(\phi_\sigma,\phi_\pi\right)=&(1-r^2)\phi_\sigma^2+(1+r^2)\phi_\pi^2-2\int_{-\pi}^{\pi}\frac{d^3k}{(2\pi)^3}\log X_0
-\frac{\beta}{3}(1+r^2)\sum_j\left[\int_{-\pi}^{\pi}\frac{d^3k}{(2\pi)^3}\frac{\sin^2 k_j}{X_0}\right]^2\\
&-\frac{\beta}{3}(1+r^2)\left[\frac{1+r^2}{2}\phi_\pi\int_{-\pi}^{\pi}\frac{d^3k}{(2\pi)^3}\frac{\sum_j\cos k_j}{X_0}\right]^2
-\frac{\beta}{3}(1-r^2)\left[\int_{-\pi}^{\pi}\frac{d^3k}{(2\pi)^3}\frac{\tilde{m}(\bm{k})+r}{X_0}\sum_j\cos k_j\right]^2+\mathcal{O}(\beta^2), \label{F_eff-final}
\end{aligned}
\end{equation}
where we have defined $X_0= {\sum}_j\sin^2k_j+\left[\tilde{m}(\bm{k})+r\right]^2+\frac{(1+r^2)^2}{4}\phi_\pi^2$.
\end{widetext}
\section{Numerical Results}
At first, we found that the value of $\phi_\pi$ is zero at the stationary point for any set of $(r,m_0)$.
Hence in the following, we set $\phi_\sigma=-\sigma$ and $\phi_\pi=0$ in Eq. (\ref{Feff-SCL}) to calculate the value of the chiral condensate $\sigma$.
The term $i\bar{\psi}\gamma_5\psi$ is odd under both time-reversal and inversion.
Therefore, this means that the phase with spontaneously broken time-reversal and inversion symmetries does not arise in the strong coupling (electron correlation) limit.
A mean-field study of Wilson fermions with the short-range interaction from the weak coupling\cite{Sekine2012}  and a lattice strong coupling expansion study of the Kane-Mele model on a honeycomb lattice\cite{Araki2013} suggest the existence of this phase.
Such a phase, "Aoki phase" (where parity and flavor symmetry are spotaneously broken) has also confirmed in the lattice QCD with Wilson fermions\cite{Aoki1984,Aoki1986,Sharpe1998}.
We mention the main difference between this analysis and lattice QCD
except for the gauge group as follows.
Our effective model has only temporal (timelike) link variablies in contrast with lattice QCD.
Spatial link variables are absent, like in the case of free fermions. 
Parity-flavor symmetry is not spontaneously broken in free fermions. 
This is one of the reasons why the parity broken phase does not appear in this analysis.

\begin{figure}[!t]
\begin{center}
\includegraphics[width=\columnwidth,clip]{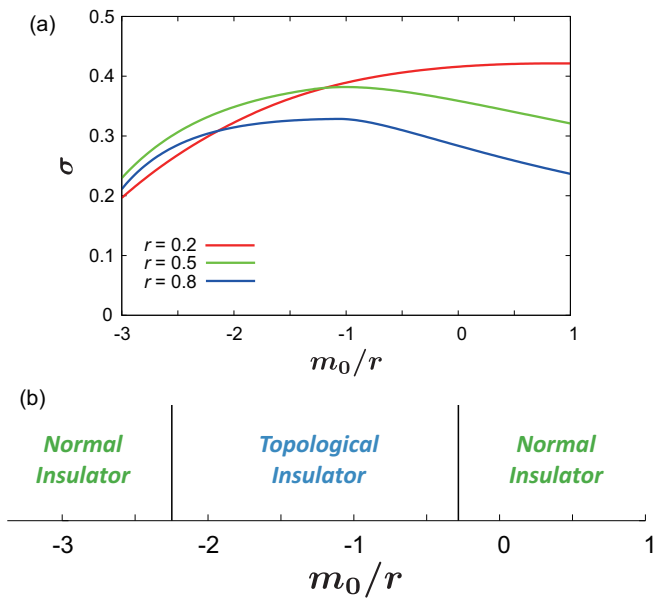}
\caption{(Color online) (a) $m_0$-dependence of the chiral condensate $\sigma$ in the strong coupling limit ($\beta=0$).
(b) Phase diagram with $r=0.5$ in the strong coupling limit. The phase boundaries are determined by the condition $m_{\rm eff}=0$ or $m_{\rm eff}=-2r$.}\label{fig1}
\end{center}
\end{figure}

The $m_0$-dependence of the chiral condensate $\sigma$ is shown in Fig. \ref{fig1}(a).
The value of $\sigma$ is expected to be quantitatively correct, based on the fact that the result of a strong coupling expansion study in graphene\cite{Araki2010,Araki2011} is in good agreement with that of lattice Monte Carlo studies\cite{Hands2008,Drut2009,Drut2009a,Armour2010}.
As mentioned above, in the noninteracting limit (i.e. at $\beta=\infty$), the system with $0>m_0>-2r$ ($m_0>0$) is identified as a topological (normal) insulator.
The chiral condensate is equivalent to a correction to the bare mass.
Hence it is natural to define the effective mass in Eq. (\ref{m_eff}):
\begin{equation}
\begin{aligned}
m_{\rm{eff}}=m_0+(1-r^2)\sigma/2.
\end{aligned}
\end{equation}
The phase diagram with $r=0.5$ in the strong coupling limit calculated by the $Z_2$ invariant (Eq. (\ref{Z2invariant})) is shown in Fig \ref{fig1}(b).
In the strong coupling limit, the system with $0>m_{\rm eff}>-2r$ ($m_{\rm eff}>0$) is identified as a topological (normal) insulator.
From this phase diagram, we see that the effect of the long-range Coulomb interaction is to shift the region of the topological insulator phase.
This result doesn't contradict that of a mean-field analysis from the weak coupling\cite{Sekine2012}.

\begin{figure}[!t]
\begin{center}
\includegraphics[width=0.83\columnwidth,clip]{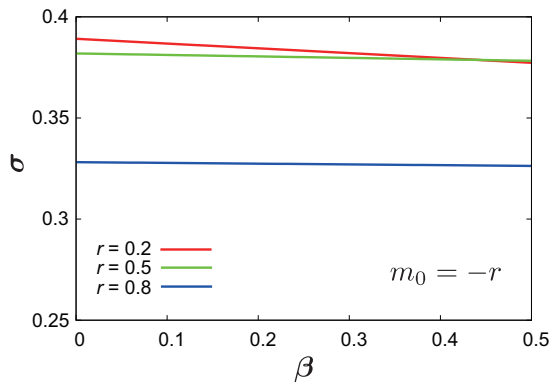}
\caption{(Color online) $\beta$-dependence of the chiral condensate $\sigma$ at $m_0=-r$.}\label{fig2}
\end{center}
\end{figure}

The $\beta$-dependence of the chiral condensate $\sigma$ is shown in Fig. \ref{fig2}.
We see that $\sigma$ is a monotonically decreasing function of the coupling strength $\beta$.
This behavior is consistent with a mean-field analysis from the weak coupling\cite{Sekine2012}.
Our result shows that the mass gap remains finite, in contrast to the mean-field analysis in which the mass gap becomes infinity in the strong coupling limit.
We see also that as $r$ becomes smaller, the rate of decrease of $\sigma$ becomes notable.
Namely, as the original mass of doublers becomes smaller, the energy gap of the system becomes smaller, as is understood intuitively.

From Fig \ref{fig2}, it is concluded that the gapped phases (normal or topological insulator phases) are stable in the strong coupling region.
This contrasts with the result of the strong coupling expansion in graphene\cite{Araki2010,Araki2011}.
In graphene, the rate of decrease of $\sigma$ from $\beta=0$ to $\beta=0.5$ is about 60\%\cite{Araki2011}, whereas that of our model is about 3\% at $r=0.2$.
Namely, in our model, the topological insulator phase survives in the strong coupling limit, although graphene undergoes the semimetal-insulator transition in the strong coupling region.

\section{Discussion and Summary}
\begin{figure}[!t]
\begin{center}
\includegraphics[width=0.90\columnwidth,clip]{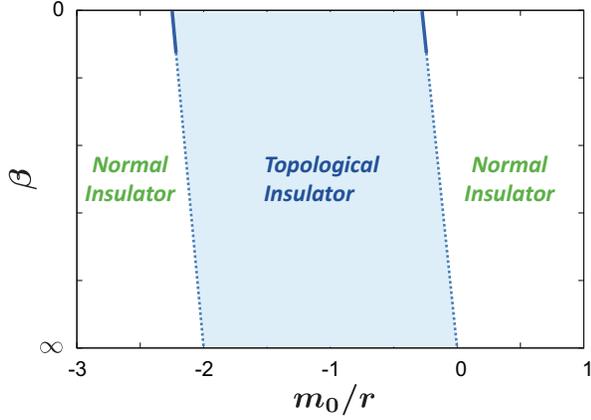}
\caption{(Color online) A possible phase diagram of the Wilson fermions interacting via the long-range Coulomb interaction ($r=0.5$).}\label{fig3}
\end{center}
\end{figure}

So far we have obtained the value of the effective mass up to of the order of the coupling strength $\beta$.
We can connect the phase boundary in the noninteracting limit and that in the strong coupling region.
A possible phase diagram of the Wilson fermions interacting via the long-range Coulomb interaction is shown in Fig. \ref{fig3}.
A similar behavior of the phase boundary between the topological insulator phase and the normal insulator phase have been obtained in a mean-field analysis of the Wilson fermions with the short-range interaction\cite{Sekine2012}.
One might wonder why the topological insulator phase survives at infinite coupling.
If the interaction is short-range, i.e., Hubbard-like, the antiferromagnetic phase is considered to be dominant.
However, in the present case, the interaction is pure $1/r$ Coulomb interaction.
This difference may affect the phase structure.
A lattice strong coupling expansion study of the Kane-Mele model on a honeycomb lattice shows a similar result that when spin-orbit coupling is sufficiently strong, the topological insulator phase survives in the strong coupling limit.

To summarize, we have studied the strong electron correlation effect in a 3D topological insulator which effective Hamiltonian can be described by the Wilson fermions.
Based on the U(1) lattice gauge theory, we have performed the strong coupling expansion.
It was found that the effect of long-range Coulomb interaction corresponds to the renormalization of the bare mass.
The values of the chiral condensate, which is regarded as a correction to the bare mass in the strong coupling limit, are expected to be correct quantitatively.
The behavior of the chiral condensate in our model is similar to that of the lattice QCD with Wilson fermions.
The phase where time-reversal and inversion symmetries are spontaneously broken ("Aoki phase") was not found in the strong coupling region, in contrast to the case of lattice QCD.
It was also found that the gapped phase is stable in the strong coupling region.
This suggests that the topological insulator phase survives in the strong coupling limit.
In this study, the bulk property of a 3D topological insulator was examined.
It will be interesting to examine the strong correlation effect in the surface Dirac fermions.

\begin{acknowledgments}
T. Z. N. is thankful to H. Iida, D. Satow, and S. Gongyo for fruitful discussions.
Y. A. is thankful to T. Kimura for valuable discussions.
This work was supported by the Grants-in-Aid for Scientific Research (No. 24740211 and No. 10J03314) from the Ministry of Education, Culture, Sports, Science and Technology, Japan (MEXT).
A. S. is supported by the global COE program "Weaving Science Web beyond Particle-Matter Hierarchy" from MEXT.
T. Z. N. is supported by the Fellowship for Young Scientists (No. 22-3314) from Japan Society for the Promotion of Science (JSPS) and the global COE program "The Next Generation of Physics, Spun from Universality and Emergence" from MEXT.
Y. A. is supported by JSPS Postdoctoral Fellowship for Research Abroad (No.25.56).
\end{acknowledgments}

\appendix
\begin{widetext}
\section{Detailed calculation of the effective potential up to $\bm{\mathcal{O}(\beta)}$}
In this appendix, we show the detailed calculation of the effective Potential up to $\mathcal{O}(\beta)$.
The terms which consist of only auxiliary fields in Eqs. (\ref{Delta S_1}) and (\ref{Delta S_2}) are obtained as
\begin{equation}
\begin{aligned}
S_{\alpha\beta}S'_{\alpha\beta}&=\left\langle B^T\right\rangle_{\beta\alpha}\left\langle A\right\rangle_{\beta\alpha}=-\left\langle V^-_{n+0,j}P^-_0\right\rangle_{\alpha\beta}\left\langle V^+_{n,j}P^+_0\right\rangle_{\beta\alpha}\\
&= - \mathrm{tr} \left[ \langle V_{n,j}^- \rangle P_0^- \langle V_{n,j}^+ \rangle P_0^+ \right]\\
&=(1-r^2) v_s^- v_s^+ +(1+r^2) v_p^- v_p^+ +(1-r^2) v_{v0}^- v_{v0}^+ - (1+r^2) \sum_l v_{vl}^- v_{vl}^+ -(1+r^2) v_{a0}^- v_{a0}^+ + (1-r^2) \sum_l v_{al}^- v_{al}^+,
\end{aligned}
\end{equation}
and
\begin{equation}
\begin{aligned}
T_{\alpha\beta}T'_{\alpha\beta}&=\left\langle B^T\right\rangle_{\beta\alpha}\left\langle A\right\rangle_{\beta\alpha}=-\left\langle V^+_{n+0,j}P^-_0\right\rangle_{\alpha\beta}\left\langle V^-_{n,j}P^+_0\right\rangle_{\beta\alpha}\\
&= - \mathrm{tr} \left[ \langle V_{n,j}^+ \rangle P_0^- \langle V_{n,j}^- \rangle P_0^+ \right]\\
&=(1-r^2) v_s^- v_s^+ +(1+r^2) v_p^- v_p^+ +(1-r^2) v_{v0}^- v_{v0}^+ - (1+r^2) \sum_l v_{vl}^- v_{vl}^+ -(1+r^2) v_{a0}^- v_{a0}^+ + (1-r^2) \sum_l v_{al}^- v_{al}^+.
\end{aligned}
\end{equation}

The fermionic terms in Eq. (\ref{Delta S_1}) are obtained as
\begin{equation}
\begin{aligned}
-A_{\alpha \beta} S_{\beta\alpha} 
=& \left( V_{n,j}^+ P_0^{+T} \right)_{\alpha \beta} \langle V_{n+\hat{0},j}^- P_0^{-T} \rangle_{\beta \alpha}
= \bar{\psi}_n \left( P_0^+ \langle V_{n,j}^- \rangle P_0^- \right)^T \psi_{n+\hat{j}}
\end{aligned}
\end{equation}
with
\begin{equation}
\begin{aligned}
P_0^+ \langle V_{n,j}^- \rangle P_0^-
=&
\displaystyle \frac {1}{4} \Biggl[
(r^2-1) v_s^- + i \gamma_5 \left\{ (r^2+1) v_p^- - 2r v_{a0}^- \right\}+ \gamma_0 \cdot (r^2-1) v_{v0}^- 
\\
&+ \sum_k \gamma_k \cdot (r^2+1) v_{vk}^- + i \gamma_5 \gamma_0 \left\{ - 2 r v_p^- + (r^2+1) v_{a0}^- \right\} + \sum_k i \gamma_5 \gamma_k \cdot (r^2-1) v_{ak}^- 
\Biggr],
\end{aligned}
\end{equation}
and
\begin{equation}
\begin{aligned}
-B_{\alpha \beta}^T S'_{\beta\alpha} 
= \left( V_{n+\hat{0},j}^- P_0^{-T} \right)_{\beta \alpha} \langle V_{n,j}^+ P_0^{+T} \rangle_{\alpha \beta}
= \bar{\psi}_{n+\hat{j}} \left( P_0^- \langle V_{n,j}^+ \rangle P_0^+ \right)^T \psi_n
\end{aligned}
\end{equation}
with
\begin{equation}
\begin{aligned}
P_0^- \langle V_{n,j}^+ \rangle P_0^+
=&
\displaystyle \frac {1}{4} \Biggl[
(r^2-1) v_s^+ + i \gamma_5 \left\{ (r^2+1) v_p^+ + 2r v_{a0}^+ \right\}+ \gamma_0 \cdot (r^2-1) v_{v0}^+ 
\\
&+ \sum_k \gamma_k \cdot (r^2+1) v_{vk}^+ + i \gamma_5 \gamma_0 \left\{ 2 r v_p^+ + (r^2+1) v_{a0}^+ \right\} + \sum_k i \gamma_5 \gamma_k \cdot (r^2-1) v_{ak}^+ 
\Biggr].
\end{aligned}
\end{equation}

Similary, the fermionic terms in Eq. (\ref{Delta S_2}) are obtained as
\begin{equation}
\begin{aligned}
-A_{\alpha \beta} T_{\beta\alpha} 
= \left( V_{n,j}^- P_0^{+T} \right)_{\alpha \beta} \langle V_{n+\hat{0},j}^+ P_0^{-T} \rangle_{\beta \alpha} 
= \bar{\psi}_{n+\hat{j}} \left( P_0^+ \langle V_{n,j}^+ \rangle P_0^- \right)^T \psi_{n} 
\end{aligned}
\end{equation}
with
\begin{equation}
\begin{aligned}
P_0^+ \langle V_{n,j}^+ \rangle P_0^-
=&
\displaystyle \frac {1}{4} \Biggl[
(r^2-1) v_s^+ + i \gamma_5 \left\{ (r^2+1) v_p^+ - 2r v_{a0}^+ \right\}+ \gamma_0 \cdot (r^2-1) v_{v0}^+ 
\\
&+ \sum_k \gamma_k \cdot (r^2+1) v_{vk}^+ + i \gamma_5 \gamma_0 \left\{ - 2 r v_p^+ + (r^2+1) v_{a0}^+ \right\} + \sum_k i \gamma_5 \gamma_k \cdot (r^2-1) v_{ak}^+ 
\Biggr],
\end{aligned}
\end{equation}
and
\begin{equation}
\begin{aligned}
-B_{\alpha \beta}^T T'_{\beta\alpha} 
= \left( V_{n+\hat{0},j}^+ P_0^{-T} \right)_{\beta \alpha} \langle V_{n,j}^- P_0^{+T} \rangle_{\alpha \beta} 
= \bar{\psi}_{n} \left( P_0^- \langle V_{n,j}^- \rangle P_0^+ \right)^T \psi_{n+\hat{j}}
\end{aligned}
\end{equation}
with
\begin{equation}
\begin{aligned}
P_0^- \langle V_{n,j}^- \rangle P_0^+
=&
\displaystyle \frac {1}{4} \Biggl[
(r^2-1) v_s^- + i \gamma_5 \left\{ (r^2+1) v_p^- + 2r v_{a0}^- \right\}+ \gamma_0 \cdot (r^2-1) v_{v0}^- \\
&+ \sum_k \gamma_k \cdot (r^2+1) v_{vk}^- + i \gamma_5 \gamma_0 \left\{ 2 r v_p^- + (r^2+1) v_{a0}^- \right\} + \sum_k i \gamma_5 \gamma_k \cdot (r^2-1) v_{ak}^- 
\Biggr].
\end{aligned}
\end{equation}

Then doing the Fourier transform and combining Eqs. (\ref{S_eff_SCL}) and (\ref{Delta S}), we get the effective action up to $\mathcal{O}(\beta)$ with auxiliary fields:
\begin{equation}
\begin{aligned}
S_{\mathrm{eff}}=S_{\mathrm{eff}}^{\mathrm{aux}}\left(\phi_\sigma,\phi_\pi, v^\pm_s, v^\pm_p, v^\pm_{v\mu}, v^\pm_{a\mu}\right)+\sum_k \bar{\psi}_k\mathcal{M}\left(\bm{k};\phi_\sigma,\phi_\pi, v^\pm_s, v^\pm_p, v^\pm_{v\mu}, v^\pm_{a\mu}\right)\psi_k,
\end{aligned}
\end{equation}
where
\begin{equation}
\begin{aligned}
S_{\mathrm{eff}}^{\mathrm{aux}}=&N_sN_\tau\left\{ (1-r^2)\phi_\sigma^2+(1+r^2)\phi_\pi^2+\beta\sum_{j}\left[(1-r^2) v_s^- v_s^+ +(1+r^2) v_p^- v_p^+ +(1-r^2) v_{v0}^- v_{v0}^+\rule{0pt}{4ex}\right.\right.\\
&\left.\left.- (1+r^2) \sum_l v_{vl}^- v_{vl}^+ -(1+r^2) v_{a0}^- v_{a0}^+ + (1-r^2) \sum_l v_{al}^- v_{al}^+ \right] \right\},
\end{aligned}
\end{equation}
and
\begin{equation}
\begin{aligned}
\mathcal{M}
=&
m_0 + r \left( 4 - \sum_j \cos k_j \right) + \displaystyle \frac{\sigma}{2} \left( r^2 - 1 \right) \cos \theta 
 + \sum_j\frac{\beta}{4} (r^2-1) \left[ v_s^- e^{ik_j} + v_s^+ e^{-ik_j} \right]+ i \gamma_5^T \frac{\sigma}{2} \left( r^2 + 1 \right) \sin \theta\\
& + i \gamma_5^T \sum_j \displaystyle \frac{\beta}{4} (r^2+1) \left[ v_p^- e^{ik_j} + v_p^+ e^{-ik_j} \right]+ \gamma_0^T \sum_j \frac{\beta}{4} (r^2-1) \left[ v_{v0}^- e^{ik_j} + v_{v0}^+ e^{-ik_j} \right]+ \sum_j i \gamma_j \sin k_j \\
& + \sum_{j,l} \gamma_k^T \frac{\beta}{4} (r^2+1) \left[ v_{vl}^- e^{ik_j} + v_{vl}^+ e^{-ik_j} \right]+ i \sum_j\left( \gamma_5 \gamma_0 \right)^T\frac{\beta}{4} (r^2+1) \left[ v_{a0}^- e^{ik_j} + v_{a0}^+ e^{-ik_j} \right]\\
& + \sum_{j,l} i \left( \gamma_5 \gamma_l \right)^T\frac{\beta}{4} (r^2-1) \left[ v_{al}^- e^{ik_j} + v_{al}^+ e^{-ik_j} \right].
\end{aligned}
\end{equation}
We replace the auxiliary fields to make the calculation easier as follows:
\begin{equation}
\begin{aligned}
&v_s^{\mp} = S_1 \mp i S_2,\ \ \ v_p^{\mp} = \mp i P_1 + P_2,\ \ \ v_{v0}^{\mp} = i V_{0,1} \pm V_{0,2},\\ &v_{vk}^{\mp} = \mp V_{k,1} -i V_{k,2},\ \ \ v_{a0}^{\mp} = \mp A_{0,1} -i A_{0,2},\ \ \ v_{ak}^{\mp} = -i A_{k,1} \mp A_{k,2}.\;\label{new_auxiliaryfields}
\end{aligned}
\end{equation}
Using the formula $\mathrm{det}\mathcal{M}=\sqrt{\mathrm{det}(\mathcal{M}\mathcal{M}^\dag)}$, we obtain
\begin{equation}
\begin{aligned}
\mathrm{det}\mathcal{M}=\left[
\mathcal{M}_s^2
+ \mathcal{M}_p^2
 + \mathcal{M}_{v0}^2
+ \sum_j \mathcal{M}_{vj}^2
 + \mathcal{M}_{a0}^2 
 + \sum_j \mathcal{M}_{aj}^2 \right]^2,
\end{aligned}
\end{equation}
where we have defined $\mathcal{M}$ as
\begin{equation}
\begin{aligned}
\mathcal{M}
= \mathcal{M}_s
+ i \gamma_5^T \mathcal{M}_p
 + i \gamma_0^T \mathcal{M}_{v0}
+ \sum_j i \gamma_j \mathcal{M}_{vj} + \left( \gamma_5 \gamma_0 \right)^T \mathcal{M}_{a0} 
 + \sum_j \left( \gamma_5 \gamma_j \right)^T \mathcal{M}_{aj}.
\end{aligned}
\end{equation}

The final form of the effective potential (Eq. (\ref{F_eff-final})) is obtained by the stationary conditions:
\begin{equation}
\begin{aligned}
\frac{\partial \mathcal{F}_{\mathrm{eff}}}{\partial S}=\frac{\partial \mathcal{F}_{\mathrm{eff}}}{\partial P}=\frac{\partial \mathcal{F}_{\mathrm{eff}}}{\partial V}=\frac{\partial \mathcal{F}_{\mathrm{eff}}}{\partial A}=0,
\end{aligned}
\end{equation}
where $S$, $P$, $V$ and $A$ are the auxiliary fields defined in Eq. (\ref{new_auxiliaryfields}).

\end{widetext}

\end{document}